\documentstyle[preprint,pra,aps,eqsecnum]{revtex}

\def\ket#1{|#1\rangle}

\def\n{\vec n}
\def\m{\vec m}

\def\tr{{\rm tr}}

\def\dOn{d\Omega_{\tilde n}\,}
\def\dOm{d\Omega_{\tilde m}\,}
\def\dOa{d\Omega_{\tilde a}\,}
\def\dOb{d\Omega_{\tilde b}\,}

\begin{document}
\tighten

\title{Classical model for bulk-ensemble \\ NMR quantum computation}

\author{
R\"udiger Schack\thanks{E-mail: r.schack@rhbnc.ac.uk} 
 $^{^{\hbox{\tiny (a,b)}}}$
and
Carlton M. Caves\thanks{E-mail: caves@tangelo.phys.unm.edu} 
 $^{^{\hbox{\tiny (a)}}}$}
\address{
$^{\hbox{\tiny (a)}}$Center for Advanced Studies, Department of Physics
and Astronomy, \\
University of New Mexico, Albuquerque, NM~87131--1156, USA \\
$^{\hbox{\tiny(b)}}$Department of Mathematics, Royal Holloway, \\ 
University of London, Egham, Surrey TW20 0EX, UK
}

\date{\today}

\maketitle

\begin{abstract}
  We present a classical model for bulk-ensemble NMR quantum computation: 
  the quantum state of the NMR sample is described by a probability 
  distribution over the orientations of classical tops, and quantum gates 
  are described by classical transition probabilities.  All NMR quantum 
  computing experiments performed so far with three quantum bits can be 
  accounted for in this classical model.  After a few entangling gates, 
  the classical model suffers an exponential decrease of the measured 
  signal, whereas there is no corresponding decrease in the quantum 
  description. We suggest that for small numbers of quantum bits, the 
  quantum nature of NMR quantum computation lies in the ability to avoid 
  an exponential signal decrease.
\end{abstract}

\section{Introduction}

The original proposals \cite{Cory1997,Gershenfeld1997} for quantum computing 
using high-temperature, liquid-state nuclear magnetic resonance (NMR) 
sparked an explosion of interest.  The resulting work on NMR 
quantum-information processing has attracted a great deal of attention.  
In addition to a long list of publications \cite{Cory1997,Gershenfeld1997,%
Warren1997,Nielsen1998a,Cory1998a,Laflamme1998a,Linden1998a,Cory1998c,%
Jones1998d,Chuang1998c,Chuang1998a,Chuang1998b,Knill1998a,Cory1998b,%
Jones1998a,Jones1998c,Jones1998b,Berman1998,Knill98,Schulman1998},
there have been numerous news reports, especially on the recent NMR 
experiments on quantum error correction \cite{Cory1998a} and quantum 
teleportation \cite{Nielsen1998a}.  

The fundamental information-processing elements in NMR are two-level 
nuclear spins, called quantum bits, or {\it qubits} for short, which are 
bound together in a single molecule.  A liquid NMR sample contains a 
macroscopic number of molecules, each of which functions as an independent 
information-processing system.  The molecules are initially in thermal 
equilibrium at high temperature; the nuclear spins are only weakly 
polarized along the direction of a strong applied magnetic field.  NMR 
techniques cannot control the quantum states of individual molecules; 
instead, all the molecules in the sample are manipulated in parallel.  
Moreover, NMR readout techniques are sensitive to the average magnetization 
of the entire sample.  For these reasons the use of high-temperature, 
liquid-state NMR techniques for quantum-information processing is called 
{\it bulk-ensemble quantum computation}.  Clever techniques have been 
devised to map the coherent quantum-mechanical manipulations required 
for quantum computation to this situation in which one neither controls 
nor makes measurements on individual molecules 
\cite{Cory1997,Gershenfeld1997,Cory1998c,Chuang1998a,Knill1998a}.  

This paper deals exclusively with bulk-ensemble quantum-information 
processing, not with proposals (see, for example, \cite{Kane1997}) for 
using nuclear spins in situations, as in solid-state systems, where they
can be highly polarized.  Although the paper is phrased in terms of
and aimed at NMR quantum computation, the analysis is not specific to
NMR and could be applied to any realization of bulk-ensemble quantum 
computation. 

Bulk-ensemble NMR quantum computation is in principle scalable to many 
qubits \cite{Schulman1998}, but it is unclear whether it will ever lead 
to useful quantum computations involving more than a few dozen qubits 
\cite{Cory1997,Warren1997}.  Nevertheless, the experiments performed so 
far, which use only a few qubits, are cited as important demonstrations 
of the principles of quantum-information processing and quantum computation.  
There has, however, been persistent skepticism about whether these 
experiments demonstrate genuine quantum-information processing.  Recently 
these doubts have been brought into sharper focus \cite{DiVincenzo1998b} 
with the realization that all quantum states used in present NMR experiments 
are {\it separable} \cite{Braunstein1999a}, i.e., unentangled.  Entanglement 
is generally thought to be an essential feature of quantum computation 
\cite{Ekert1998,Zurekreply,onequbit}.  

In this paper we explore the ``quantumness'' of bulk-ensemble quantum 
computation.  There being no general definition of what it means to be 
doing genuine quantum-information processing, we begin by proposing three 
different criteria, which are investigated in the remainder of the paper.

\begin{itemize}

\item[$\bullet$]
According to the first criterion, an $N$-qubit NMR experiment does 
not demonstrate genuine quantum-information processing if all quantum
states that occur during the experiment are {\it separable\/} (unentangled), 
i.e., are equivalent to classical ensembles of spinning tops.  The motivation 
for this criterion is that for separable states, the statistics of any 
measurement performed on the NMR sample can be described in the language
of classical probabilities for spin orientations.  With respect to this 
criterion, NMR experiments for up to about 15 qubits do not demonstrate 
genuine quantum-information processing \cite{Braunstein1999a}.  There is, 
however, an immediate objection to this criterion: even if all states 
involved in the experiment are classical, the transformations between the 
states might be essentially quantum mechanical, not having any description 
in classical language.

\item[$\bullet$]
Our second criterion addresses this objection by stating that an 
$N$-qubit NMR experiment does not demonstrate genuine quantum-information 
processing if we can construct an {\it overall classical model\/} for 
the experiment, by which we mean that in addition to the experiment's 
satisfying the first criterion, the transformation $T$ that maps the 
initial quantum state to the final quantum state can be described by 
classical transition probabilities.  The transition probabilities are 
required to be fully specified by $T$; in particular, they must be 
independent of the initial and final states.  We formulate an overall 
classical model below, which shows that, with respect to this second 
criterion, NMR experiments for up to about 6 qubits do not demonstrate 
genuine quantum-information processing.  An objection to this second 
criterion is that an overall classical model does not constitute a valid 
model for a quantum computation, because the complexity of the overall
transition probabilities might increase exponentially with problem size; 
according to this objection, computational complexity is at the heart of 
the question of whether a given computation is quantum mechanical.

\item[$\bullet$]
Our third criterion incorporates this objection as follows.  It 
states that an $N$-qubit NMR experiment does not demonstrate genuine quantum
information processing if we can construct a {\it gate-by-gate classical
model}, by which we mean that in addition to the first criterion being 
satisfied, any gate $U$ implemented in the experiment can be described by 
classical transition probabilities, which, as before, are fully specified 
by $U$ and which, apart from trivial contributions for the qubits not 
involved in the gate, are independent of the total number of qubits
\cite{Bcaveat}.  We devise a gate-by-gate classical model below.  
It shows that all 3-qubit quantum computing 
experiments performed so far 
\cite{Nielsen1998a,Cory1998a,Laflamme1998a,Linden1998a,Cory1998c} do 
not demonstrate genuine quantum-information processing according to 
this third criterion.

\end{itemize}

In the NMR literature, transformations are often represented by diagrams
that depict spins as arrows evolving on spheres; see, e.g., \cite{NMRdiagrams}
or the ``effective fields picture'' in 
\cite{Cory1998b}, where a diagram representing a controlled-NOT gate is 
given.  This diagram illustrates what is new in our approach.  The diagram 
assumes that the sample is split into two subpopulations, according to 
whether the control qubit is up or down.  Thus it provides a classical 
description of the gate for initial conditions corresponding to 
computational basis states of the control qubit, but not for arbitrary, 
possibly entangled initial states.  Consequently, it does not give a 
classical model of the controlled-NOT gate in the sense of our third 
criterion.

We emphasize that our gate-by-gate classical model is not the same as
a simulation of a quantum algorithm on a classical computer.  First, in 
our classical {\it model}, measurement statistics are determined by 
probabilities for classical spin orientations; in a classical 
{\it simulation\/}, where one calculates the quantum amplitudes in, 
say, the spin-up--spin-down, or computational basis, one cannot go from 
the amplitudes to measurement statistics using classical probability logic.  
This means that our classical model provides a hidden-variable description 
of the experiment \cite{Bcaveat}, but a classical simulation does not.  
Second, the computational complexity of a classical simulation increases 
exponentially with the number of gates \cite{Ekert1998}, whereas the 
complexity of our transition probabilities for an entangling gate increases 
only linearly with the number of qubits.

Equally important is a difference between our gate-by-gate classical model 
and the quantum description of an NMR experiment.  In our classical model, 
the strength of the magnetization signal decreases exponentially with the 
number of entangling gates.  Though this decrease can be staved off for 
a few entangling steps, it cannot be put off indefinitely.  Such a decrease 
is absent from the quantum description.  Thus it is straightforward to 
design experiments that cannot be described by our classical model.  
Two-qubit experiments that fall into this class are \cite{Jones1998d} 
and probably also \cite{Chuang1998c}, although in the latter case this 
is not clear from the published data.  In order to prove conclusively, 
however, that a bulk-ensemble NMR experiment demonstrates genuine 
quantum-information processing in the sense of the third criterion given 
above, one must rule out {\it all\/} gate-by-gate classical models, not 
just the particular model given in this paper.  

The paper is organized as follows.  In Sec.~\ref{sec:background}, we 
give a brief description of bulk-ensemble NMR quantum computation, and
in Sec.~\ref{sec:sep}, we summarize the proof given in \cite{Braunstein1999a} 
that the states used in high-temperature NMR experiments are classical up 
to about 15 qubits in the sense of the first criterion above.  In 
Sec.~\ref{sec:transprobs}, we define the transition probabilities that
are used in our overall and gate-by-gate classical models.  Used directly 
in a gate-by-gate classical model, these transition probabilities give rise 
to a signal decrease by a constant factor at each entangling multi-qubit 
operation.  Section~\ref{sec:model} shows how to modify the gate-by-gate 
model so as to avoid this signal decrease for a small number of entangling 
steps.  In Sec.~\ref{sec:discussion} we consider the implications of our 
results for bulk-ensemble NMR quantum-information processing.

\section{NMR quantum-information processing} 
\label{sec:background}

All NMR quantum computing experiments performed so far 
\cite{Nielsen1998a,Cory1998a,Laflamme1998a,Linden1998a,Cory1998c,Jones1998d,%
Chuang1998c,Chuang1998a,Chuang1998b,Knill1998a,Cory1998b,Jones1998a,Jones1998c}
work according to the following principles.  The state of the sample is 
described by a density operator $\rho$ for the $N$ spins (qubits) in each 
molecule.  The molecules are prepared in an initial state
\begin{equation}
  \rho = (1-\epsilon)M + \epsilon\rho_1 \;,
  \label{eq:rhoepsilon}
\end{equation}
where $M=I/2^N$ is the maximally mixed density operator for $N$ qubits ($I$
is the unit operator) and $\rho_1$ is a density operator, usually chosen 
to be the projector onto the computational basis state $\ket{0\ldots0}$.  
The parameter $\epsilon$ scales like
\begin{equation}
  \epsilon = {\alpha N\over2^N} \;.
  \label{eq:epsscaling}
\end{equation}
Here $\alpha=h\bar\nu/2kT$, where $\bar\nu$ is the average resonant frequency
of the active spins in the strong magnetic field, determines the polarization
of the sample.  If $\rho_1$ is a pure state, $\rho$ is called a pseudopure 
state \cite{Cory1997}.  Procedures for synthesizing a pseudopure state from 
an initial thermal density operator are described in 
\cite{Gershenfeld1997,Cory1998c,Chuang1998a,Knill1998a}.  

For typical molecules in present magnetic-field strengths, the average
resonant frequency is roughly $\bar\nu\sim300\,$MHz, which at room temperature, 
gives $\alpha\sim2\times10^{-5}$.  It is difficult to know the actual value 
of $\alpha$ and $\epsilon$ in the experiments, because the polarization is 
not absolutely calibrated.  In evaluating the experiments, we adopt the 
scaling of Eq.~(\ref{eq:epsscaling}), with a conservative value of 
$\alpha=2\times10^{-6}$, which is meant to take into account inefficiencies 
in the experiments, especially loss of polarization involved in the synthesis 
of a pseudopure state.

One interpretation of $\epsilon$ is that it specifies the fraction of 
molecules in the sample that occupy the desired initial state $\rho_1$.  
This interpretation is not unique, and it is not mandated or even preferred 
by quantum mechanics.  It is typically promoted to the level of a physical 
fact by advocates of NMR quantum computation, yet it becomes a physical 
fact only if one actually prepares a fraction of the molecules in a 
particular state \cite{Vidal1999}, a situation that does not apply 
to a high-temperature NMR experiment. The freedom to interpret $\rho$ in 
terms of other ensembles underlies the conclusions of this paper.

After synthesis of the desired initial state, a sequence of radio-frequency 
pulses, alternating with continuous evolution, is applied to the sample. 
We first consider the case where the evolution is described by a unitary 
transformation $U$.  Applying the unitary operator $U$ to $\rho$ results 
in the state
\begin{equation}
 \rho^{\rm out} \equiv U\rho\, U^\dagger 
     = (1-\epsilon)M + \epsilon U\rho_1 U^\dagger
  \equiv  (1-\epsilon)M + \epsilon \rho_1^{\rm out}   \;. 
\end{equation}
The totally mixed part of the state is unaffected by the unitary 
transformation.  The output state retains the form (\ref{eq:rhoepsilon})
with the same value of $\epsilon$, but---and this is the essence of the
bulk-ensemble paradigm for quantum computation---$\rho_1$ undergoes the 
desired unitary transformation.  

The same conclusions hold for transformations implemented using gradient 
pulses together with diffusion in the sample \cite{Cory1998c}; these are 
equivalent to mixtures of unitary transformations, as in 
\begin{equation}
 \rho^{\rm out} \equiv \sum_l p_lU_l\rho\,U_l^\dagger = (1-\epsilon)M 
      + \epsilon \sum_l p_lU_l\rho_1 U_l^\dagger  
   \equiv (1-\epsilon)M + \epsilon \rho_1^{\rm out} \;,
  \label{eq:unitarymix}
\end{equation}
where the $p_l\ge0$ are probabilities.  We assume that decoherence also 
preserves the form of the density operator, leaving $\epsilon$ unchanged.  
This assumption is certainly justified when decoherence is simulated by 
unitary transformations, as in the gradient pulses used to simulate 
decoherence in~\cite{Cory1998a}.  It is also true for naturally occurring 
decoherence for times short compared to the characteristic time 
for relaxation to thermal equilibrium.  These are the times of interest 
for quantum-information processing.  

The last step in any NMR experiment is the readout.  By applying 
radio-frequency pulses and then measuring the transverse magnetization 
of the sample, one can determine all expectation values of the form
\begin{equation}
  \tr\Bigl(\rho^{\rm out}  
       \sigma_{\beta_1}\otimes\cdots\otimes\sigma_{\beta_N}\Bigr)
 = \epsilon\,\tr\Bigl( \rho_1^{\rm out}
              \sigma_{\beta_1}\otimes\cdots\otimes\sigma_{\beta_N}\Bigr) \;.
\label{eq:expect}
\end{equation}
The tensor product in this expression includes one operator for each
spin; $\sigma_{\beta_k}$ denotes the unit operator $I$ for the $k$th spin 
if $\beta_k=0$, and it denotes a Pauli matrix if $\beta_k=1,2$, or 3.  In 
writing Eq.~(\ref{eq:expect}) and all such expectation values in this 
paper, it is assumed that there is at least one Pauli matrix in the 
tensor product (not all the $\beta$'s are zero).  The maximally mixed 
part of the density operator does not contribute to the measured expectation 
values, which are determined by the state $\rho_1^{\rm out}$ that undergoes 
the desired evolution.  The parameter $\epsilon$ appears naturally as a 
measure of the strength of the magnetization signal (or of the 
signal-to-noise ratio).  

\section{Separability of states used in NMR} 
\label{sec:sep}

Our conclusions rest on the 
freedom to write states of the form~(\ref{eq:rhoepsilon}) in terms of 
probability distributions over spin orientations for $N$ classical tops.  
We review one such representation \cite{Braunstein1999a}, the foundation 
for our work, which provides a classical description for all states of 
the form~(\ref{eq:rhoepsilon}) provided that
\begin{equation}
  \epsilon \le \eta \equiv {1\over1+2^{2N-1}} \;.
\end{equation}
If we assume that $\alpha=2\times10^{-6}$ in Eq.~(\ref{eq:epsscaling}), this 
inequality holds for $N<16$ qubits. 

The argument is straightforward \cite{Braunstein1999a}.  The most general
pure product state of $N$ qubits has the form
\begin{equation}
P_1(\tilde n)\equiv P_1(\vec n_1,\ldots,\vec n_N)
= {1\over2^N}(I+\vec n_1\cdot\vec\sigma)\otimes\cdots
\otimes(I+\vec n_N\cdot\vec\sigma) \;,
\label{eq:pureprod}
\end{equation}
This state can be interpreted as $N$ classical tops pointing in the 
directions given by the unit vectors $\vec n_1,\ldots,\vec n_N$, denoted
collectively by $\tilde n$.  Since the operators $P_1(\tilde n)$ form an 
overcomplete basis in the space of linear operators acting on $N$ qubits, 
any $N$-qubit density operator $\rho$ can be expanded as
\begin{equation}
  \rho = \int \dOn w^\rho(\tilde n) P_1(\tilde n) \;,
\label{eq:P1expansion}
\end{equation}
where $\dOn\equiv d\Omega_{\vec n_1}\cdots d\Omega_{\vec n_N}$.  The 
overcompleteness means that the expansion coefficients are not unique.  
One choice is \cite{Braunstein1999a}
\begin{equation}
w^\rho(\tilde n) \equiv \tr\Bigl(\rho Q_1(\tilde n)\Bigr) \;,
\label{eq:wrho}
\end{equation}
where the operators $Q_1(\tilde n)$ are defined by
\begin{equation}
Q_1(\tilde n)\equiv{1\over(4\pi)^N}
(I+3\vec n_1\cdot\vec\sigma)\otimes\cdots
\otimes(I+3\vec n_N\cdot\vec\sigma) \;.
\end{equation}
The expansion coefficients $w^\rho(\tilde n)$ can be positive or negative, 
but they obey the bound
\begin{equation}
 w^\rho(\tilde n) \ge 
\pmatrix{\hbox{smallest}\cr\hbox{eigenvalue}\cr\hbox{of $Q_1(\tilde n)$}}
=-{2^{2N-1}\over(4\pi)^N} \;.
\end{equation}
When $w^\rho(\tilde n)$ is everywhere nonnegative, it can be interpreted as
a classical probability distribution for the $N$ spins to point in the 
directions $\vec n_1,\ldots,\vec n_N$.  Since the maximally mixed density 
operator $M$ has probability density $w^M(\tilde n)=1/(4\pi)^N$, it follows 
that for a density operator $\rho$ of the form (\ref{eq:rhoepsilon}) with 
$\epsilon\le\eta$, 
\begin{equation}
  w^\rho(\tilde n) =
{1-\epsilon\over(4\pi)^N}+\epsilon w^{\rho_1}(\tilde n)\ge 
{1-\epsilon/\eta\over(4\pi)^N} 
\ge0 \;.
\label{eq:nonneg}
\end{equation}

A density operator for a joint system is {\it separable\/} if it can be 
written as a nonnegative linear combination of product density operators.  
When $w^\rho(\tilde n)$ is everywhere nonnegative, the 
expansion~(\ref{eq:P1expansion}) provides a separable representation for 
$\rho$.  Related work on the separability of states near the maximally
mixed state can be found in \cite{Vidal1999} and \cite{Zyczkowski1998}.
Separable states have no quantum entanglement \cite{entanglediscuss}.  
The importance of separability in this paper is that a separable state 
of $N$ qubits can be interpreted in terms of an ensemble of classical tops,
because the expectation values~(\ref{eq:expect}) have the standard form for 
an ensemble with probability distribution $w^\rho(\tilde n)$: 
\begin{equation}
\tr\Bigl(\rho\,\sigma_{\beta_1}\otimes\cdots\otimes\sigma_{\beta_N}\Bigr)
= 
\int\dOn w^\rho(\tilde n)\,
\underbrace{\tr\Bigl(
P_1(\tilde n)\sigma_{\beta_1}\otimes\cdots\otimes\sigma_{\beta_N}
\Bigr)}_{\displaystyle{\mbox{}=(n_1)_{\beta_1}\cdots(n_N)_{\beta_N}}} 
\;. 
\end{equation}
In this expression, $(n_j)_{\beta_j}=1$ if $\beta_j=0$, and $(n_j)_{\beta_j}$ 
is a Cartesian component of the vector $\n_j$ if $\beta_j=1,2$, or 3.

If one adopts our first criterion, none of the NMR experiments performed 
to date has done any genuine quantum-information processing, since 
they all use 2 or 3 qubits.  Yet the fact that all states---initial, 
intermediate, and final---that occur in a given NMR experiment are 
equivalent to ensembles of classical tops does not mean, by itself, that 
there is a classical model for the entire experiment.  To see this, 
consider the following na{\"\i}ve attempt to describe a unitary 
transformation $U$ by classical transition probabilities:
\begin{equation}
  w^{U\rho\,U^\dagger}(\tilde n) 
= \tr\Bigl(U\rho\,U^\dagger Q_1(\tilde n)\Bigr) 
= \int\dOm w^\rho(\tilde m) 
  \underbrace{\tr\Bigl(UP_1(\tilde m) U^\dagger Q_1(\tilde n)\Bigr)}_{
  \displaystyle{\mbox{}\equiv t_U(\tilde n|\tilde m)}} 
\;.
\label{eq:tU}
\end{equation}
The transition function $t_U(\tilde n|\tilde m)$ is not a transition 
probability because it assumes negative values; for example, for the 
trivial case of the identity transformation, $U=I$, we have
\begin{equation}
t_I(\tilde n|\tilde m)
=\tr\Bigl(P_1(\tilde m)Q_1(\tilde n)\Bigr)
={1\over(4\pi)^N}\prod_{j=1}^N(1+3\m_j\cdot\n_j)
\;. 
\end{equation}

The first criterion is the right one, however, for judging claims of having 
produced entangled states using bulk-ensemble NMR.\ For example, as pointed 
out in \cite{Braunstein1999a}, the claim of Laflamme {\it et al.} 
\cite{Laflamme1998a} to have created a 3-qubit Greenberger-Horne-Zeilinger 
(GHZ) entangled state is incorrect.  Despite the authors' assertion, ``We 
describe the creation of a Greenberger-Horne-Zeilinger state of the form 
$(|000\rangle+|111\rangle)/\sqrt2$ (three maximally entangled quantum bits) 
using nuclear magnetic resonance. \ \dots\ We have thus extended the space 
of entangled quantum states explored systematically to three quantum 
bits\ \dots\ ,'' no entanglement was created in that experiment; the 
statistics of any measurement performed on the purported GHZ state could 
have been explained in terms of classical correlations contained in 
$w^\rho(\tilde n)$.  Similar conclusions apply to the experiments in which
Chuang {\it et al.}\ \cite{Chuang1998a} and Cory {\it et al.}\ 
\cite{Cory1998c} claim to have created 2-qubit Einstein-Podolsky-Rosen 
(EPR) entangled states.  The states created in these experiment were 
unentangled, though Chuang {\it et al.}\  write, ``As an application 
of the controlled-NOT gate, we used it in a simple quantum circuit to 
create entangled states from the thermal mixture. \ \dots\ We have 
experimentally confirmed this (nonclassical) behavior, and the signature 
of the entanglement---a purely {\it non-classical\/} effect---is the 
strong reverse diagonal measured in the density matrix,'' and Cory 
{\it et al.} refer to applying the ``spin-coherence XOR gate to a one-spin 
superposition to create an entangled state.''

Indeed, the first criterion can be applied to any NMR experiment that 
claims to have manipulated quantum {\it states\/} in a particular way.  
For example, in describing the NMR version of quantum teleportation
\cite{Nielsen1998a}, a 3-qubit experiment, the authors assert, 
``Quantum-mechanical systems have information processing capabilities 
that are not possible with classical devices.  One example is quantum 
teleportation, in which the quantum state of a system is transported from 
one location to another without moving through the intervening space. \ 
\dots\ Here we report an experimental implementation of full quantum 
teleportation over inter-atomic distances using liquid-state nuclear 
magnetic resonance.''  This claim cannot be supported, because the 
quantum state at all stages of the experiment could be interpreted in 
terms of classical correlations among spin directions.  What the experiment 
achieved was a reshuffling of these classical correlations.  Nonetheless, 
as noted above, this conclusion does not imply that there is a classical 
model for the entire experiment.  

\section{Transition probabilities for quantum gates} 
\label{sec:transprobs}

The essence of a classical model lies in the construction of nonnegative 
transition probabilities to describe unitary transformations.  We first 
consider product unitaries, i.e., unitary transformations of the form 
$V=V_1\otimes\cdots\otimes V_N$, and then tackle the tougher task of 
entangling unitaries, i.e., unitary transformations that are not product 
unitaries.

The expansion coefficients for the output state are given by
\begin{equation}
  w^{V\rho V^\dagger} (\tilde n)
= \tr\Bigl(V\rho V^\dagger Q_1(\tilde n)\Bigr)
= \tr\Bigl(\rho V^\dagger Q_1(\tilde n)V \Bigr) 
\;.
\end{equation}
For a product unitary $V$, the unitary transform of $Q_1(\tilde n)$ 
factors into a product of transformations for each qubit.  If we introduce 
the 3-dimensional rotation operator $R_k$ (an orthogonal transformation) 
corresponding to the unitary operator $V_k$, i.e., 
$V_k^\dagger\vec\sigma V_k=R_k\vec\sigma$, and use the fact that
$\n_k\cdot R_k\vec\sigma=R_k^{-1}\n_k\cdot\vec\sigma$, then the unitary
transform of $Q_1(\tilde n)$ assumes the form
\begin{equation}
 V^\dagger Q_1(\tilde n)V 
=Q_1(R_1^{-1}\n_1,\ldots,R_N^{-1}\n_N)=
Q_1(\tilde R^{-1}\tilde n)
\;.
\end{equation}
Here $\tilde R$ stands for the rotations on all the qubits.  Now the output 
expansion coefficients assume the form
\begin{equation}
w^{V\rho V^\dagger} (\tilde n)
= w^\rho(\tilde R^{-1}\tilde n) 
= w^\rho(R_1^{-1}\n_1,\ldots,R_N^{-1}\n_N)
\;.
\end{equation}
Not surprisingly, the action of the product unitary $V$ is equivalent to 
rotating each of the 
classical tops.  The corresponding nonnegative transition probabilities are 
$w(\tilde n|\tilde m) = \delta(\tilde R^{-1}\tilde n - \tilde m)$.  This 
result shows that, for product unitaries, our transition probabilities are 
equivalent to the simple classical diagrams often used in the NMR literature 
\cite{NMRdiagrams}.

Before proceeding to entangling unitaries, we describe how decoherence is 
incorporated into our classical models.  In some experiments decoherence 
is simulated as a mixture of product unitary transformations; an example 
is the use of gradient pulses in~\cite{Cory1998a}.  Naturally occurring 
decoherence processes act independently on the various qubits; on time 
scales short compared to the thermal relaxation time, they preserve the 
maximally mixed density operator.  Any decoherence process that satisfies 
these two properties can be described as a mixture of product unitaries 
\cite{Landau1993}.  Thus decoherence can be handled in our classical models 
as a mixture of rotations of the qubits. 

To describe general, entangling unitaries, we need to introduce some 
additional notation.  For an arbitrary density operator $\rho$ and 
for $0<\theta\le1$, we define
\begin{equation}
  \rho_\theta\equiv(1-\theta)M + \theta\rho \;.
\end{equation}
Furthermore, we define the states
\begin{equation}
  P_\theta(\tilde n)\equiv(1-\theta)M + \theta P_1(\tilde n) 
\end{equation}
and associated operators
\begin{equation}
  Q_\theta(\tilde n)\equiv 
      {1-\theta^{-1}\over(4\pi)^N}\,I + \theta^{-1} Q_1(\tilde n) \;.
\end{equation}
For a general unitary operator $U$, we define the transition probabilities
\begin{equation}
   w_U(\tilde n|\tilde m)\equiv 
      \tr\Bigl(UP_\eta(\tilde m) U^\dagger Q_1(\tilde n)\Bigr) =
      {1-\eta\over(4\pi)^N}+\eta t_U(\tilde n|\tilde m)
         \ge0 \;,
\label{eq:wcond}
\end{equation}
where $t_U(\tilde n|\tilde m)$ is defined in Eq.~(\ref{eq:tU}).
The nonnegativity of these transition probabilities follows from the 
argument leading to Eq.~(\ref{eq:nonneg}), since $UP_\eta(\tilde m)U^\dagger$ 
is a state of the form~(\ref{eq:rhoepsilon}) with $\epsilon=\eta$.  

It is straightforward to write down the transition probabilities for 
elementary entangling gates.  As an illustration, we compute them for the 
controlled-phase gate,
\begin{eqnarray}
U=C_{ij}
&=&|0\rangle\langle0|\otimes1+|1\rangle\langle1|\otimes\sigma_3\nonumber\\
&=&{1\over2}(1+\sigma_3)\otimes1+
{1\over2}(1-\sigma_3)\otimes\sigma_3
\;,
\end{eqnarray}
acting on qubits $i$ and $j$.  We obtain
\begin{eqnarray}
w_{C_{ij}}(\tilde n|\tilde m)
&=& \tr\Bigl( C_{ij} P_\eta(\tilde m) C_{ij} Q_1(\tilde n) \Bigr) \nonumber\\
&=& {1\over(4\pi)^N} \Biggl(
     1-\eta + \eta \prod_{l\ne i,j}(1+3\m_l\cdot\n_l) \nonumber \\
&\mbox{}&\phantom{x}
     \times\Bigl([1+3(m_i)_3(n_i)_3][1+3(m_j)_3(n_j)_3]
           +9(\m_i\times\n_i)_3 (\m_j\times\n_j)_3   \nonumber \\
&\mbox{}&\phantom{x\times}
         + 3 [(m_i)_3+3(n_i)_3](\m_{j\perp}\cdot\n_{j\perp})
         + 3 [(m_j)_3+3(n_j)_3](\m_{i\perp}\cdot\n_{i\perp})
   \Bigr)\Biggl) \;, \nonumber \\
\end{eqnarray}
where $\m_\perp$ is the projection of $\m$ into the 1-2 plane, obtained by
setting $m_3$ to zero.  As in this example, the qubits not involved in an 
entangling gate do appear in the transition probabilities, but in a simple, 
universal way.

Applying the transition probabilities~(\ref{eq:wcond}) to an input ensemble 
$w_{\rm in}\equiv w^\rho(\tilde n)$, we obtain a classical output 
ensemble
\begin{eqnarray}
    w_{\rm out}(\tilde n) &\equiv& \int\dOm w_U(\tilde n|\tilde m) 
        w_{\rm in}(\tilde m)  \nonumber\\
&=& \tr\Biggl(U
  \underbrace{\biggl[\int\dOm w^\rho(\tilde m)P_\eta(\tilde m)\biggr]}_{
  \displaystyle{\mbox{}=\rho_\eta}} 
              U^\dagger Q_1(\tilde n)\Biggr) \nonumber\\
&=& w^{U\rho_\eta U^\dagger}(\tilde n) \nonumber \\
&=& {1-\eta\over(4\pi)^N}+\eta w^{U\rho\,U^\dagger}(\tilde n)\;.
\label{eq:wout}
\end{eqnarray}
Unlike the quantum output ensemble $w^{U\rho\,U^\dagger}(\tilde n)$, the 
classical output ensemble suffers an increase in the fraction of molecules 
that are maximally mixed.  This means that the magnetization signal produced 
by the classical output ensemble is a factor of $\eta$ smaller than in the 
quantum description:
\begin{eqnarray}
\int\dOn w_{\rm out}(\tilde n)(n_1)_{\beta_1}\cdots(n_N)_{\beta_N}
&=& \tr\Bigl(  U\rho_\eta U^\dagger 
      \sigma_{\beta_1}\otimes\cdots\otimes\sigma_{\beta_N}\Bigr) \nonumber\\
&=& \eta\,\tr\Bigl(  U\rho\,U^\dagger 
      \sigma_{\beta_1}\otimes\cdots\otimes\sigma_{\beta_N}\Bigr) \;.
\end{eqnarray}

If we use the transition probabilities~(\ref{eq:wcond}) to describe 
transitions of the spin directions in a gate-by-gate classical model of 
an NMR quantum computation, the magnetization signal of the model loses 
a factor of $\eta$ at each entangling gate.  Though such a model gives
a satisfactory account of an NMR experiment in which one ignores the 
strength of the signal, we can formulate a better model that avoids the 
decay for a few entangling gates.

\section{Gate-by-gate classical model} 
\label{sec:model}

The key idea in constructing an improved model is to introduce auxiliary, 
``hidden'' spins 
$\vec a_1,\ldots,\vec a_N\equiv\tilde a$, one for each qubit.  We also 
need a ``counter index'' $k$, which increments by 1 at each entangling 
gate.  In the improved model, we can avoid the decay of the signal for 
$K$ entangling gates, where $K$ is the largest integer such that
\begin{equation}
  \epsilon \le \eta^{K+1} \;.
\end{equation}
We assume that $\epsilon\le\eta^2$ to ensure that $K\ge1$.  It is useful 
to introduce the function
\begin{equation}
\eta_k\equiv\cases{
       \eta^{K-k}\;,&$0\le k< K$,\cr
       1\;,         &$k\ge K$.}
\end{equation}

Between the $k$th and $(k+1)$th entangling gates ($k\ge0$), we represent 
a density operator $\rho$ by the expansion coefficients
\begin{equation}
w_k^\rho(\tilde a)\equiv  
\tr\Bigl(\rho Q_{\eta_k}(\tilde a)\Bigr) \;.
\label{eq:wkrho}
\end{equation}
It is easy to show that these expansion coefficients are related to the 
original coefficients~(\ref{eq:wrho}) by
\begin{equation}
w_k^\rho(\tilde a)={1-\eta_k^{-1}\over(4\pi)^N}+\eta_k^{-1} w^\rho(\tilde a)
\label{wrelation}
\end{equation}
and that $\rho$ can be expanded as
\begin{equation}
\rho = \int\dOa w_k^\rho(\tilde a) P_{\eta_k}(\tilde a) \;. 
\label{eq:Pkexpansion}
\end{equation}
For density operators of the form~(\ref{eq:rhoepsilon}), Eqs.~(\ref{wrelation})
and~(\ref{eq:nonneg}) imply that
\begin{equation}
w_k^\rho(\tilde a)\ge{1-\epsilon/\eta^{K+1}\over(4\pi)^N}\ge0
\end{equation}
for all $k\ge0$.  Throughout the experiment, the expansion coefficients 
$w_k^\rho(\tilde a)$ can be interpreted as probability distributions.

In our improved classical model, the gate operations manipulate the hidden 
spins; the effect of a gate shows up in the observable spins $\tilde n$ 
through the correlations of $\tilde n$ with $\tilde a$.  Between the $k$th 
and $(k+1)$th entangling gates, this correlation is described by the joint 
distribution
\begin{equation}
w_k^\rho(\tilde n,\tilde a)\equiv 
q_k(\tilde n|\tilde a) w_k^\rho(\tilde a) \;,
\end{equation}
where 
\begin{equation}
q_k(\tilde n|\tilde a)\equiv
{1-\eta_k\over(4\pi)^N}+\eta_k\delta(\tilde n-\tilde a)
\end{equation}
is a conditional probability distribution.  The marginal distribution for 
the observable spin variables,
\begin{equation}
\int\dOa w_k^\rho(\tilde n,\tilde a) =
{1-\eta_k\over(4\pi)^N}+\eta_k w_k^\rho(\tilde n)=
w^\rho(\tilde n)\;,
\end{equation}
is just the right mixture of $w_k^\rho(\tilde n)$ with the uniform 
distribution to produce the distribution $w^\rho(\tilde n)$ that 
gives the measured expectation values.
 
We now define transition probabilities for quantum gates, distinguishing
as before between product unitaries and entangling unitaries.  For a product 
unitary $V$, we find, by an argument analogous to the previous one, that
\begin{eqnarray}
  w_k^{V\rho V^\dagger} (\tilde n,\tilde a) 
= w_k^\rho(\tilde n,\tilde R^{-1}\tilde a) \;.
\end{eqnarray}
The effect of $V$ is to rotate the hidden spins.  Decoherence is handled, 
as previously, by mixtures of product unitaries.

To deal with an entangling unitary transformation $U$, we first note that
the transition probabilities $w_U(\tilde a|\tilde b)$ defined in 
Eq.~(\ref{eq:wcond}) can be expressed as
\begin{equation}
  w_U(\tilde a|\tilde b) = 
        \tr\Bigl(U P_{\eta\theta}(\tilde b) U^\dagger 
        Q_\theta(\tilde a)\Bigr) \ge0 \;.
\end{equation}
Using this result, together with Eq.~(\ref{eq:Pkexpansion}), we find
\begin{eqnarray}
  \int\dOb w_U(\tilde a|\tilde b) w_k^\rho(\tilde b)
   &=&\tr\Biggl( U
     \underbrace{\biggl[
        \int\dOb P_{\eta\eta_{k+1}}(\tilde b)w_k^\rho(\tilde b)\biggr]}_{
        \displaystyle{\mbox{}=
            \cases{
               \rho\;,     &$0\le k<K$\cr
               \rho_\eta\;,&$k\ge K$}
                     }}
     U^\dagger Q_{\eta_{k+1}}(\tilde a)\Biggr) \nonumber\\
  &=&\cases{ 
        w_{k+1}^{U\rho\,U^\dagger}(\tilde a)\;,     &$0\le k<K$,\cr
        w_{k+1}^{U\rho_\eta U^\dagger}(\tilde a)\;,&$k\ge K$.}
\label{eq:wu}
\end{eqnarray}
To describe the effect of the $(k+1)$th entangling gate, we use 
transition probabilities
\begin{equation}
  w_{k+1}^U(\tilde n,\tilde a|\tilde m,\tilde b)
     \equiv q_{k+1}(\tilde n|\tilde a) w_U(\tilde a|\tilde b) \;,
\end{equation}
which take the input ensemble $w_{\rm in}\equiv w^\rho_k(\tilde n,\tilde a)$
to an output ensemble
\begin{eqnarray}
    w_{\rm out}(\tilde n,\tilde a) 
&=& \int\dOm\dOb w_{k+1}^U(\tilde n,\tilde a|\tilde m,\tilde b)
        w_{\rm in}(\tilde m,\tilde b)    \nonumber\\
&=& q_{k+1}(\tilde n|\tilde a)
    \int\dOb w_U(\tilde a|\tilde b) w_k^\rho(\tilde b)   \nonumber\\
&=& \cases{
        w_{k+1}^{U\rho\,U^\dagger}(\tilde n,\tilde a)\;,&$0\le k<K$,\cr
        w_{k+1}^{U\rho_\eta U^\dagger}(\tilde n,\tilde a)\;,&$k\ge K$.}
\end{eqnarray}

We have now constructed a gate-by-gate classical model in which there is
no loss of magnetization signal for the first $K$ entangling gates.  How
this works can be summarized as follows.  The initial distribution 
$w_0^\rho(\tilde a)$ of the hidden spin variables is chosen to have as
little contribution from the uniform distribution as is consistent 
with nonnegativity.  At each entangling gate, the ensemble of hidden spins 
suffers an increase in the proportion of molecules that are maximally mixed, 
but the observable spins retain the statistics of the quantum description 
by becoming more tightly correlated with the hidden spins.  After $K$ 
entangling gates, the observable and hidden spins become $\delta$-correlated; 
thus from the $(K+1)$th entangling gate on, the signal decreases by a 
factor of $\eta$ at each entangling gate.

\section{Discussion} 
\label{sec:discussion}

Consider a bulk-ensemble NMR experiment whose parameter
$\epsilon$ satisfies the condition $\epsilon\le\eta^2$ that underlies
the argument in Sec.~\ref{sec:model}.  If we assume $\alpha=2\times10^{-6}$ 
in Eq.~(\ref{eq:epsscaling}), this condition is fulfilled for $N\le6$ qubits.
The model constructed in Sec.~\ref{sec:model} provides an overall 
classical model in the sense of our second criterion: the unitary 
transformation $U$ that maps the initial state to the final state is
described in the model, with no loss of magnetization signal, by the 
transition probabilities $w_1^U (\tilde n,\tilde a|\tilde m,\tilde b)$; 
mixtures of such transition probabilities are used to incorporate 
decoherence.  The predictions of the model for the signal derived from 
the output state are identical to the quantum predictions.  Since all 
NMR experiments to date involve 2 or 3 qubits, this overall classical 
model applies to all such experiments.

We turn now to the implications of the gate-by-gate classical model 
constructed in Sec.~\ref{sec:model}.  If we assume again that 
$\alpha=2\times10^{-6}$, in a 2-qubit experiment the model proceeds 
through $K=5$ entangling gates with no signal loss, and in a 3-qubit 
experiment, through $K=3$ entangling gates.  We illustrate the implications 
by considering a particular 3-qubit experiment, the NMR version of quantum 
teleportation \cite{Nielsen1998a}.  After preparation of a pseudopure 
state by the gradient-pulse technique \cite{Cory1998c}, the teleportation 
experiment consisted of four operations: (i)~an entangling 2-qubit 
gate, (ii)~a 2-qubit mapping of the Bell basis to the computational 
basis, (iii)~decoherence of two qubits in the computational basis, and 
(iv)~a conditional 3-qubit unitary.  Our model incorporates the decoherence 
step as a mixture of classical rotations of the hidden spins.  Since the 
model can account for $K=3$ entangling gates without loss of signal, it 
provides a gate-by-gate classical model whose predictions are identical 
to the quantum predictions.  This experiment therefore does not demonstrate 
genuine quantum-information processing in the sense of our third criterion.
For a larger value of $\alpha$, when $K=2$, our model predicts a drop 
in signal by a factor of $\eta=1/33$ at the third entangling step.  
Nevertheless, even when $K=2$, the classical model can still account for 
the teleportation experiment if the first two 2-qubit gates are ``compiled'' 
into a single 3-qubit gate. 

Similar conclusions apply to the other 3-qubit experiments performed 
to date \cite{Cory1998a,Laflamme1998a,Linden1998a,Cory1998c} and to most 
of the 2-qubit experiments 
\cite{Chuang1998a,Chuang1998b,Knill1998a,Cory1998b,Cory1998c,Jones1998a,%
Jones1998c}. The 2-qubit experiment described in \cite{Jones1998d} ran 
through 15 entangling gates with an approximately exponential signal-to-noise 
decrease that is, however, much slower than the factor of $\eta=1/9$ per 
entangling gate predicted by our model after the first $K=5$ steps.  The 
2-qubit experiment reported in \cite{Chuang1998c} implemented up to 14 
entangling gates, but the absence of signal-to-noise data makes comparison 
with our model difficult.

This paper, together with \cite{Braunstein1999a}, begins the task of 
establishing standards for assessing the quantumness of bulk-ensemble 
NMR quantum computation \cite{Schulmancomment,quotes}.  Our gate-by-gate 
classical model erects a hurdle in the way of NMR quantum-information 
processing.  ``Testing'' our classical model is not the point, for no one 
would contend that it describes the physics that underlies an NMR experiment.  
The point is that experiments that fail to clear the hurdle can be explained 
in classical language and thus do no genuine quantum-information processing 
according to our third criterion.

It will be easy for NMR experimenters to clear our hurdle, by measuring 
the signal-to-noise ratio in experiments involving many entangling 
gates, like the experiment reported in \cite{Jones1998d}.  Yet jumping 
over our hurdle is probably not sufficient to provide a warranty of genuine 
quantum-information processing, for that would require showing that 
a given experiment is inconsistent with {\it all\/} gate-by-gate classical
models, not just the model formulated in this paper.  There are reasons 
for believing that our classical model is not optimal, the most cogent
of which is that the model's transition probabilities for an entangling 
gate transform qubits that are unaffected by the gate.  Avenues for 
improving the model include the following: (i)~investigating more efficient 
representations both for quantum states near the maximally mixed density 
operator and for the transition probabilities between states, (ii)~seeking 
gate representations that are more efficient when fewer qubits are involved 
in the gate, and (iii)~addressing the extent to which one is allowed to 
``compile'' successive entangling unitaries into a single operation 
(compiling is routine in NMR experiments as a method for reducing the 
length of a computation).  

We conjecture, however, that no matter how efficient the gate representations
are made, an ultimate signal decrease is an unavoidable consequence of any 
attempt to describe entangling unitaries classically, even when the unitaries 
act only on separable states.  Indeed, more interesting than our results 
would be a demonstration of this conjecture.  Should the conjecture prove to 
be correct, one could conclude that the quantumness of NMR quantum computation, 
for small numbers of qubits, lies in the ability to avoid any signal decrease.
More broadly, we speculate that the power of quantum-information processing 
comes not from entanglement itself, but rather from the information-processing
capabilities of entangling unitaries \cite{onequbit}.

\acknowledgments
Thanks to S.~L. Braunstein for drawing our attention to the question of 
the quantumness of NMR quantum computation, to R.~Jozsa, M.~Grassl, and 
D.~Janzing for useful comments, and to C.~A. Fuchs for a careful reading of
and suggestions for improving the manuscript.  This work was supported in 
part by the US Office of Naval Research (Grant No.~N00014-93-0-0116) and 
by the UK Engineering and Physical Sciences Research Council.


\begin{thebibliography}{10}
\bibitem{onequbit} Knill and Laflamme \cite{Knill98} consider a model for 
ensemble computation using input states in which one qubit is in a pure 
state and all the other qubits are in a maximally mixed state.  They note
that although the apparent power of quantum computers ``is frequently 
attributed to `quantum parallelism,' interference phenomena derived from 
the superposition principle, and the ability to prepare and control pure
states according to the Schr{\"o}dinger equation,'' their results suggest
that ``the usual reasons given for why quantum computation appears to be
so powerful may have to be revised.''
\bibitem{Bcaveat}
It is likely that there are ``nonlocal'' hidden-variable models for $N$ 
qubits which reproduce all the predictions of quantum mechanics, but 
which violate the assumptions of our third criterion, in that the 
transition probabilities for a gate would depend on the initial state
or would include qubits not involved in a gate in a nontrivial way.
\end{thebibliography}
\end{document}